\documentclass[acmtog, authorversion]{acmart}

\renewcommand\footnotetextcopyrightpermission[1]{} 

\makeatletter
\renewcommand\@formatdoi[1]{\ignorespaces}
\def\runningfoot{\def\@runningfoot{}}
\def\firstfoot{\def\@firstfoot{}}
\def\evenfoot{\def\@evenfoot{}}
\def\oddfoot{\def\@oddfoot{}}
\makeatother

\usepackage{booktabs} 
\usepackage{hyperref}
\usepackage{footmisc}
\setcitestyle{square}

\usepackage{booktabs} 
\usepackage{hyperref}
\usepackage{footmisc}
\setcitestyle{square}

\usepackage[ruled]{algorithm2e} 

\SetAlFnt{\small}
\SetAlCapFnt{\small}
\SetAlCapNameFnt{\small}
\SetAlCapHSkip{0pt}
\IncMargin{-\parindent}

\setcopyright{rightsretained}

\begin{document}
\title{How many people microwork in France?}
\subtitle{Estimating the size of a new labor force}

\author{Cl\'ement Le Ludec}
\affiliation{%
  \institution{MSH Paris Saclay}
  \city{Paris}
  \country{France}
  }
\email{leludec.clement@gmail.com}

\author{Paola Tubaro}
\affiliation{%
  \institution{National Center of Scientific Research (CNRS)}
  \city{Laboratory of Research in Informatics, Paris-Saclay}
  \country{France}
}
\email{paola.tubaro@lri.fr}

\author{Antonio A. Casilli}
\affiliation{%
 \institution{Telecom ParisTech}
 \city{Paris}
 \country{France}}
\email{antonio.casilli@telecom-paristech.fr}

\renewcommand\shortauthors{Le Ludec, C. et al}

\begin{abstract}
Microwork platforms allocate fragmented tasks to crowds of providers with remunerations as low as few cents. Instrumental to the development of today's artificial intelligence, these micro-tasks push to the extreme the logic of casualization already observed in "uberized" workers. The present article uses the results of the DiPLab study to estimate the number of people who microwork in France. We distinguish three categories of microworkers, corresponding to different modes of engagement: a group of 14,903 "very active" individuals, most of whom are present on these platforms at least once a week; a second featuring 52,337 "routine" users, more selective and present at least once a month; a third circle of 266,126 "casual" users, more heterogeneous and who alternate inactivity and various levels of work practice. Our results show that microwork is comparable to, and even larger than, the workforce of ride-sharing and delivery platforms in France. It is therefore not an anecdotal phenomenon and deserves great attention from researchers, unions and policy-makers.
\end{abstract}


\maketitle

\section{Introduction}
\label{Introduction}
The digital economy currently experiences a proliferation of "microwork" platforms, specialist services where providers perform fragmented and standardized micro-tasks to be paid on a piece-rate basis. These activities take only minutes and their reward can be as low as few cents. Often repetitive and unqualified, they consist for example in identifying or naming objects on images, transcribing invoices, translating snippets of text, moderating content (such as videos), sorting or classifying search results, responding to online surveys.

These tasks are often part of data-intensive processes and lengthy supply chains, feeding activities as varied as digitization of archives, market research, management of back-office operations and most importantly, development of artificial intelligence \cite{GraySuri2017}. Recent applications of machine learning - on which smart technologies, autonomous vehicles and virtual assistants are all based - rely on the creation and maintenance of large databases that need to be annotated, refined, labeled and more generally, augmented \cite{Porter2017}. Microwork is used, first, to prepare, categorize and qualify information for automatic learning algorithms; and second, to assess their performance and if necessary, to make corrections. Growing investments by technology companies confirm the importance of microwork in the current development of artificial intelligence. Giants such as Google and Microsoft either rely on existing platforms or create their own internal microwork markets.

This novel organization of AI-driven automation in contemporary industries does not "replace" human jobs but makes human contribution to productive processes largely inconspicuous. It supports technologies where they fail, and yet it is not a selling point that companies leverage to attract and retain large userbases. Microwork has received very little media coverage and is unfamiliar to the general public. Its invisibility is striking especially if compared to other new forms of work associated with the rise of digital platforms, such as on-demand work for urban transport applications like Uber, or for real-time delivery like Deliveroo.

Microwork platforms do not generally recruit or hire workers directly, but operate as intermediaries between client companies (usually called "requesters") who publish small tasks, and individual users (usually called "workers", "contributors", or "providers") who accept and perform the tasks. This form of digital labor pushes to the extreme the logic of workforce casualization and insecurity already noted in the context of the vast public debate and legal disputes on the status of "uberized" workers. It is therefore imperative to address this emerging phenomenon. 

Strikingly, however, the size of this phenomenon is unknown: how large is the micro-workforce? Official statistics are still ill-equipped to capture the economics of digital platforms, and the (few) surveys conducted so far have conflated microwork with other platform-mediated activities such as transport and delivery (Uber, Deliveroo). Some insight comes from platforms themselves, whose claim is that their userbase is large: if Amazon Mechanical Turk, the most widely-known microwork service, boasted 500,000 providers ("Turkers") as early as 2014\footnote{"500,000 Workers from 190 countries", cited in M. Harris, "Amazon's Mechanical Turk workers protest: 'I am a human being, not an algorithm'", \textit{The Guardian}, December 3, 2014.}, the Chinese giant Witmart alone is said to exceed 12 millions\footnote{\textit{Les Echos}, "L'actu tech en Asie : Female founders veut percer le plafond de verre", June 19, 2015.}. A widely cited 2015 World Bank report proposed a much more conservative estimate of 4.8 million people registered on microwork websites globally, just over 10\% of whom are actually active \cite{Kuek2015}.

Not only are these figures very divergent, but they are also static, no serious attempt to update them having been undertaken since 2014-15. They are also somewhat misleading insofar as they take little account of the European context. This is a serious limitation from the viewpoint of national and regional policy-making: lack of even approximate estimates makes it difficult to take action and hinders initiatives in support of local microworkers.

To fill this gap, we propose an estimate of the total number of people microworking in the specific case of France. A highly industrialized country and a pioneer in information technologies, France is less documented in the (still scant) literature on microwork, and is less present in international platforms such as Amazon Mechanical Turk, partly for language reasons. France has its own microwork platforms, the most popular of which is Wirk (formerly known as Foule Factory). Its webpage displays 50,000 contributors and it has had to close its registrations \cite{AmarViossat2016}.

To estimate the size of the microworking population in France, we draw on publicly available sources combined with the results of DiPLab ("Digital Platform Labour"), a comprehensive study of microwork in the French-speaking world\footnote{As part of DiPLab, we built an inventory of microwork platforms and mobile applications being used in France, collected online questionnaire data from Wirk microworkers, and interviewed in-depth a subset of Wirk respondents together with some clients, platform managers and other stakeholders. The data collection was conducted in Spring and Summer 2018 (see Section \ref{CaptureRecapture} for more details).}\label{DiPLab}. We identify three modes of engagement in microwork that correspond to three pools of users of these platforms. They comprise approximately 15,000 "very active" microworkers, 50,000 "routine" microworkers and 250,000 "casual" microworkers respectively. These estimates should be interpreted as orders of magnitude rather than precise values.

These relatively high figures matter to policy-making. To the extent that they exceed the number of contributors to more high-profile platforms such as Uber and Deliveroo, they call for the attention of industry and union leaders as well as public authorities.

To the best of our knowledge, ours is the first attempt to estimate the number of microworkers in a specific country (or other geographically or institutionally limited setting) across microwork platforms. It demonstrates the need to go beyond the most high-profile cases and to take into account local conditions in addition to global trends. In this sense, our results may inspire researchers interested in re-applying our methodology to other countries.
 
The remainder of this paper is organized as follows. After a brief presentation of the literature and of our selection of relevant platforms (section \ref{Literature}), we compare and contrast different methods for assessing the size of our target population (section \ref{Calculations}), and we give an interpretation of the results thus obtained (section \ref{Interpretation}). 

\section{Why is it so difficult to estimate the size of microwork?}
\label{Literature}
While a growing number of studies are exploring the topic of microwork, most of them have focused on one platform, the US-based Amazon Mechanical Turk, which is the oldest and by far the best known. The methods used to estimate the number of people working on this service can only partially be transposed to our questioning, because we are interested in a country as a whole, France, and we cannot therefore limit ourselves to a single platform (section \ref{MTurk}). We then proceed in stages, first by explaining our choice of platforms to be included in the analysis (\ref{PlatformChoice}). In the following section (\ref{Calculations}), we will present the empirical basis for our results.

\subsection{Estimating the size of Amazon Mechanical Turk micro-workforce}
\label{MTurk}
Two distinct scientific objectives have motivated existing research on the demographics of the Amazon Mechanical Turk workforce. The first concerns the working conditions, the precariousness and the level of remuneration of its users, sometimes with the aim of proposing concrete actions \cite{Paolacci2010, Ross2010}. The second purports to assuring the quality of the results of scientific surveys, questionnaires and experiments for which "Turkers" are recruited by research teams in disciplines such as psychology, marketing, and cognitive sciences. This involves ensuring access to a sufficiently large "pool" of respondents \cite{Keith2017}.

Attempts to estimate the number of people working on Mechanical Turk have arisen from these dissimilar research questions, insofar as the figure of 500,000 popularized by Amazon was to be probed. A first study showed the existence of a pool of 7,300 "rather active" unique workers at any given time \cite{Stewart2015}. In the same year, activist Kristy Milland conducted a six-week survey and counted about 30,000 workers\footnote{The study is not published but the author has made her calculations available: Kristy Milland, 150717 Preliminary results - Mapping study, July 17, 2015, URL: \url{https://docs.google.com/spreadsheets/d/1T3yP\_Jo4qELrwsE2NAPNs07L1AWmpAEr9vnhreGJ\-K0/edit\#gid=1993074859} (accessed 18 December 2018).}. \cite{Difallah2018} use Stewart's model and improve it, allowing for a longer observation time and taking into account the heterogeneous propensity of individuals to accept tasks. They show that about 100,000 people work on Jeff Bezos' platform, or one fifth of the advertised population.

Despite the presence of a sizable set of microworkers originating from India and smaller contingents from other countries, the first estimates of Amazon Mechanical Turk workforce paid particular attention to users from the United States - partly to serve the needs of scientific studies that need American participants. So far, no research intended to measure the total number of microworkers from a target country, all platforms combined. Moreover, no extant study dealt specifically with France and the French-speaking world so far. We need to move beyond these limitations to propose estimation methods for the different platforms operating in France.

\subsection{Which platforms operate in France?}
\label{PlatformChoice}

On Mechanical Turk, demand from French companies outweighs the supply of microwork from users located in France. Since 2013, the platform has been discouraging registrations outside the US and India, notably by imposing gift cards as the only means of payment. To find French microworkers, we therefore need to expand the scope of our investigation beyond the boundaries of Amazon. Recent research \cite{Berg2018,Forde2017} covering multiple countries including France, has added CrowdFlower, Microworkers, Clickworker\footnote{In addition to connecting its own registered microworkers and demanders, Clickworker also provides an entry point for UHRS (Universal Human Relevance System), Microsoft's proprietary microwork platform.}\label{UHRS} and, in the case of \cite{Berg2018}, Prolific, to Amazon Mechanical Turk. We include their selection of platforms except Prolific, a site specialized in surveys that exists only in English, and which does not seem to have users in France (there were no French respondents in the study of \cite{Berg2018}). The case of CrowdFlower is more complex because this platform has recently changed its name to Figure Eight, and access to its tasks is now through ClixSense, which is basically a portal to other platforms. It is therefore the latter that we will retain.

In addition to these international platforms, we include Wirk, a platform that exclusively recruits its microworkers in mainland France and has established itself as a key player in the market for micro-tasks in the country. As mentioned in section \ref{Introduction}, it recently changed its name from the originale Foule Factory ("Foule" meaning "Crowd" in French). We also include Ferpection, a French platform that recruits internationally. These services are potentially more attractive for French speakers than others like for example Microworkers, where English language is required even to sign up. Finally, we add Appen, an international platform open to recruitment in France, which lists tasks of the same nature as the others, but differs from them in that it organizes tasks in "projects" of slightly longer duration\footnote{This list is not exhaustive and there are other microwork websites and applications that are used in France. We exclude those that are partially outside the scope of our research: freelancing platforms such as Malt and 5euros.com only offer microwork tasks on an occasional basis, being generally more oriented towards projects requiring advanced skills (such as creating a logo or developing a website) over longer periods of time and with higher remuneration. We also exclude income-generating mobile applications such as BeMyEye and Roamler, which usually require physical presence in a given place, and for which we do not have sufficient data.}.

It is useful to characterize this selection of platforms according to their access policies (open or limited) and the presence of at least one welcome page in French (Table \ref{PlatformTypes}). With regard to the first criterion, notice that Wirk has closed registration of new members but reopens it intermittently in case of surge in the demand of micro-work by client firms. As for the second criterion, the presence of a French page seems to encourage access by new users.

\begin{table}%
\caption{Characteristics of microwork platforms included in our study}
\label{PlatformTypes}
\begin{minipage}{\columnwidth}
\begin{center}
\begin{tabular}{lll}
  \toprule
  \textbf{Platform} & \textbf{Open sign-up}   & \textbf{French page}\\
      &  & \\
\hline
  &  & \\
  Amazon Mechanical Turk & No & No\\
  Microworkers & Yes & No \\
  Clickworker & Yes & Yes \\
  ClixSense & Yes & No \\
  Wirk & No & Yes \\
  Ferpection & Yes & Yes \\
  Appen & No & Yes \\
    &  & \\
  \bottomrule
\end{tabular}
\end{center}
\bigskip\centering
\footnotesize\emph{Source:} authors' elaboration based on platforms' websites.

\end{minipage}
\end{table}%

\section{Which method to estimate the number of French microworkers?}
\label{Calculations}

Several approaches can be employed to estimate the size of the French microworking population, despite its lack of visibility in public debates as well as in official statistics. It is always tempting to take into account the official figures provided by the platforms (section \ref{Declarative}), although they tend to overestimate the real level of activity, as existing literature has already noted in the case of Amazon Mechanical Turk \cite{Stewart2015}. Building up on previous studies, we then apply a "capture-recapture" model, which leads to a lower estimate of the userbase of a single platform (\ref{CaptureRecapture}). This second approach, however, does not scale up to the entire nation’s microworking population which may be using different platforms. Therefore, we present an original method based on website audience tracking (\ref{Audience}), which does not face the same limitations. We cross our different methods to improve accuracy and obtain a meaningful range of estimates (section \ref{Crossing}) and finally, we refine our results by taking into account multi-activity of workers operating on several microwork platforms at the same time (\ref{Multi-homing}).

\subsection{The declarative method: relying on official figures from microwork platforms}
\label{Declarative}

Official figures published by microwork platforms (Table \ref{DeclarativeTab}) are sometimes the only source of information. For example, some of them are featured in a report of the French General Inspectorate of Social Affairs \cite{AmarViossat2016}. However as the same report points out, they call for caution. Platforms advertise the size of their userbase by counting signups without taking into account the actual level of activity, and can in this way artificially inflate the result. This effect is all the stronger as high numbers can impact potential fundraising or the attractiveness of the platform for business clients. It must be added that few platforms have systematic registration/deregistration management policies. They also differ in their degree of precision, some (such as Microworkers) giving very detailed figures and updating them regularly, others just providing approximations (such as Amazon, which never came back to its initial estimate of 500,000).

In general, international platforms calculate the number of their micro-workforce without breaking it down by country. The literature and platforms' own websites offer some approximate insight. On Amazon Mechanical Turk, French workers represent only a small fraction of the userbase \cite{Difallah2018}. On another international platform, Microworkers, France is not among the 10 most represented countries, and remains among those where less than 20\% of total transactions take place \cite{Hirth2011}. On the German platform Clickworker, France and other European countries (excluding Germany) account for a total of 25\% of the workforce\footnote{See \url{https://www.clickworker.com/about-us/clickworker-crowd/} consulted on 3/12/2018}. ClixSense and Appen are major international players in the microwork industry, but they do not specify the national composition of their registration databases. On the other hand, French platforms present a variety of situations. If on the one hand Wirk is only accessible to residents of France, Ferpection has users signing up not only from France but also from the United Kingdom, the United States, Ireland and other countries\footnote{See \url{http://help.ferpection.com/l/fr/les-utilisateurs/combien-y-a-t-il-dutilisateurs-dans-la-communaute} consulted on 3/12/2018}.

\begin{table}%
\caption{Official number of people registered at microworking sites, all countries combined.}
\label{DeclarativeTab}
\begin{minipage}{\columnwidth}
\begin{center}
\begin{tabular}{ll}
  \toprule
  \textbf{Platform} & \textbf{Registered users}\\
      & \\
  \hline
    & \\
  Amazon Mechanical Turk & 500,000 \\
  Microworkers & 1,215,829 \\
  Clickworker & 1,200,000 \\
  ClixSense & 7,000,000  \\
  Wirk & 50,000 \\
  Ferpection & 50,000 \\
  Appen & 1,000,000 \\
    & \\
  \textbf{Total} & \textbf{11,015,829} \\
  \bottomrule
\end{tabular}
\end{center}
\bigskip\centering
\footnotesize\emph{Source:} websites of platforms included in analysis (accessed in August 2017 for Amazon Mechanical Turk; September 2018 for all other platforms).

\end{minipage}
\end{table}%

\subsection{The ecological method: the "capture-recapture" model}
\label{CaptureRecapture}

To address this problem, \cite{Stewart2015} and \cite{Difallah2018} implemented on Mechanical Turk a "capture-recapture" approach commonly used in bio-ecology and epidemiology. They posted a micro-task for a relatively long period of time, allowing repeated participation. The technique consists in "capturing" participants a first time, identifying them as having already done the task, then seeing how many of them get "recaptured" a second time. This approach is based on two assumptions: first, the population is closed (nobody left the platform during the study) and second, all subjects have the same chances of being captured.

We were able to replicate this approach on the French leading microwork platform Wirk – the only one providing interfaces and support pages entirely in French. As part of our DiPLab study (see footnote \ref{DiPLab}), we distributed a questionnaire in the form of a paid task, administered twice with a 23-day interval: the first phase allowed us to obtain 505 responses and the second 492 responses\footnote{Calculations after cleaning and filtering of data.}. While this two-step collection was purely related to logistical issues, we saw it as an opportunity to make an estimate of the number of persons microworking for Wirk. Indeed, in the second collection, we detected 89 returning respondents - thus enabling us to apply the capture-recapture model.

As for the two assumptions required by the capture-recapture model, the first (closed population) was satisfied, as the short time interval between the two waves limited the number of those entering or exiting the platform. The so-called Lincoln-Petersen formula applied to the data thus collected, with $N$ the total number to be estimated, $n_1$ the number of users captured in the first wave, $n_2$ the number in the second wave and $m$ the number of recaptured participants, gives :

\begin{center}
$N = \frac{n_1 n_2}{m}$
\end{center}

Based on the number of individuals observed, the results of the calculation give a population of 2792 microworkers (Table \ref{CaptureTab}). This figure is certainly well below (5.6\%) the platform's official declarations, but it is consistent with the domain-specific knowledge of its managers who, answering our questions about the possible duration of the data collection, were confident that "over two months, [they] can mobilize up to 3000 people on this type of task".

However, this may be an underestimation. Firstly, it is noticeable that the result $N$ increases as the number of recaptured microworkers $m$ decreases. Thus, given the terms of our survey (we had indeed signified that participation in the survey was allowed only once), we could have expected a much smaller number of recaptured users and therefore a larger population. Secondly, the second assumption of the model (equal chance of capture) may not apply if users display an uneven level of involvement in microwork on Wirk. When facing the same issue on Amazon Mechanical Turk, \cite[p. 141]{Difallah2018} introduce a latent variable ($a$) that they interpret as propensity to participate: for example, it could reflect the fact that some people perform all the tasks available on a given platform, regardless of their appeal, difficulty or remuneration, while others are more selective. The initial formula is modified as follows\footnote{$N \approx N^*$ when the variance of $a$ is close to the mean, but $N < N^*$ otherwise.}:

\begin{center}
$N^* = \frac{n_1 n_2}{m}(1 + \frac{Var(a)}{E(a)^2}) = N(1 + \frac{Var(a)}{E(a)^2})$
\end{center}

\cite{Difallah2018} estimated the distribution of $a$ based on 28 observations over two years. In our case, their parameter setting\footnote{Difallah et al (2018, p. 142) consider that the propensity to participate follows a beta probability distribution, and that therefore the chance to recapture the same subject $n$ times follows a beta-binomial law with parameters $\alpha = 0.29$ and $\beta = 20.9$. With these values, and a sample of $S$ single observations, the population $N^*$ is equal to $\frac{S}{1-f(0|n,\alpha,\beta)}$.} would give $N* = 34,166$ (Table \ref{CaptureTab}), a value that still falls short of the official declarations of Wirk (50,000 as indicated above), but approximates it if we interpret it as an the order of magnitude.

\begin{figure}
  \includegraphics[width=6cm]{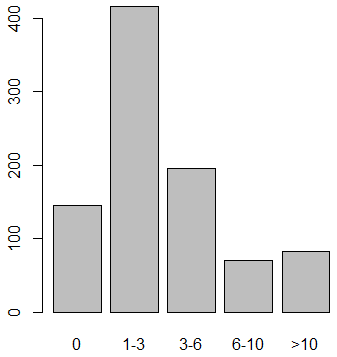}
  \Description{Number of micro-tasks carried out on Wirk in the month preceding the survey, n = 908.}
  \caption{Number of micro-tasks carried out on Wirk in the month preceding the survey, n = 908. Source: Authors' elaboration with DiPLab survey data.}
  \label{NbProjects}
\end{figure}

However, there is no guarantee that our population shares the same characteristics as people working on Mechanical Turk. We therefore propose an alternative calculation of $N^*$ using a proxy for $a$, based on one of the questions asked in our survey, touching on the number of tasks carried out by each microworker through Wirk in the previous month. The answers to this question, summarized in Figure \ref{NbProjects}, show a high diversity of participation - inconsistent with the assumption of equal chances of capture\footnote{As our variable was originally a categorical (ordered) one, we have first transformed it into a numerical variable (taking the value of the minimum of the interval in each case: 0, 1, 3, 6, 10) in order to infer a distribution from it, thus enabling calculation of its mean and variance. Alternative transformations of this variable into a numerical one give qualitatively similar results.}. Using the above formula to calculate $N*$, we obtain a population of 6531 microworkers on Wirk\footnote{Variations of this calculation, estimating distributions for each possible value in each interval, produce results of a similar order of magnitude (between 5,800 and 8,000)}.

Table \ref{CaptureTab} summarizes these results. The plain capture-recapture model (Lincoln-Petersen formula) produces an underestimation and its correction (parameters of \cite{Difallah2018}) approximates the number of official signed-up worker users (Table \ref{DeclarativeTab}). The third estimate, correcting the Lincoln-Petersen figure with a proxy from our DiPLab survey, gives an intermediate figure.

\begin{table}%
\caption{Estimate of the Wirk population using the capture-recapture method (basic formula and corrections), based on the calculations illustrated above.}
\label{CaptureTab}
\begin{minipage}{\columnwidth}
\begin{center}
\begin{tabular}{p{3.6cm}p{0.1cm}p{1.2cm}p{1.8cm}}
  \toprule
  \textbf{Method} & & \textbf{Result} & \textbf{\% Declared (50,000)} \\
      &  & \\
   \hline
     &  & \\
   Lincoln-Petersen formula & & 2,792 & 5.6\% \\
      &  & \\
   Correction with \cite{Difallah2018} parameters & & 34,166 & 68.3\% \\
         &  & \\
   Correction with a proxy from DiPLab survey & & 6,531 & 13 \% \\
     &  & \\
  \bottomrule
\end{tabular}
\end{center}
\bigskip\centering
\footnotesize\emph{Source:} authors' elaboration with DiPLab survey data.

\end{minipage}
\end{table}%

The latter estimate is more difficult to interpret based on the information provided so far. We now need to find clues to make sense of it. Also, we need to extend our analysis beyond the case of one single platform and go back to the full list of relevant microwork intermediaries that we have identified as relevant for France (section \ref{PlatformChoice}).

\subsection{The panel method: measuring the audience of microwork platforms}
\label{Audience}

The use of panels to measure the audience of these platforms, a method borrowed from the classical study of media, can provide another perspective on the population of microworkers in France. Since they receive too few visits via the M\'ediam\'etrie panel, a reference on website traffic in France, we use the alternative service SimilarWeb.com\footnote{The data used by SimilarWeb.com come from two main sources: partnerships that the site has established with Internet service providers and data that the site retrieves with an add-on on users' Internet browsers - which provides users with metrics of website traffic in exchange for their browsing data.}, which makes it possible to estimate the number of unique monthly visitors to microwork sites worldwide  (Table \ref{AudienceTab}, left column). The main difficulty in using this source is that some microwork platforms have separate interfaces for client companies (that is, their requesters) and workers. Indeed, it is the latter, namely the pages dedicated to carrying out the tasks, that we must look at. We must also exclude simple visitors from our analysis. We thus need to make the distinction explicitly and focus only on worker-only pages.

Table \ref{AudienceTab} shows that the average duration of visits is relatively long (central column). This allows us to exclude the presence of simple visitors or Internet users who are just looking for information, and to argue that these sections of the platforms are almost exclusively accessed by microworkers. The right column of Table \ref{AudienceTab} provides estimates of the average number of French unique monthly visitors per platform.

\begin{table}%
\caption{Average number of unique visitors on microwork platforms (left), average duration of visits (center) and average number of monthly unique visitors from France (right).}
\label{AudienceTab}
\begin{minipage}{\columnwidth}
\begin{center}
\begin{tabular}{p{2.8cm}p{1.5cm}p{1.3cm}p{1.5cm}}
  \toprule
   & \textbf{Average monthly} & \textbf{Average visit} & \textbf{Average monthly}\\
   \textbf{Platform} & \textbf{unique visitors} & \textbf{duration} & \textbf{unique visitors, France} \\
      &  &  & \\
   \hline
     &  &  & \\
  Worker.mturk.com & 588,976 & 34:19 & 1,250 \\
      &  &  & \\
   (Amazon Mechanical Turk) & & & \\
         &  &  & \\
   Microworkers & 174,808 & 18:22& 1,835 \\
         &  &  & \\
   Workplace.click- & 242,579 & 7:40 & 14,700 \\
   worker.com & & & \\
   (Clickworker) & & & \\
         &  &  & \\
   ClixSense & 1,083,000 & 11:05 & 20,250 \\
         &  &  & \\
   Fouleurs.com (Wirk) & 7,647 & 26:41 & 6,958 \\
         &  &  & \\
   Ferpection & 28,064 & 7:10 & 8,116 \\
         &  &  & \\
   Appen & 260,699 & 5:04 & 9,645 \\
         &  &  & \\
   & & \textbf{Total} & \textbf{62,754} \\
   
  \bottomrule
\end{tabular}
\end{center}
\bigskip\centering
\footnotesize\emph{Source:} Authors' elaboration based on information from SimilarWeb.com, accessed in September 2018. Figures are based on websites' audience tracking over the period July-September 2018.

\end{minipage}
\end{table}%

Let's first comment on the result for Wirk. 6,958 unique microkworkers access it on average every month - which is very close to the figure of 6,531 that results from our corrected capture-recapture calculation detailed in section \ref{CaptureRecapture}. This order of magnitude seems to correspond well to the size of the active population on this platform over a month, as it could be measured in the summer of 2018. It should also be noted that almost all of the visits originate from France, in accordance with the platform's policy. The residual cases of people signing in from outside France essentially concern French microworkers temporarily connecting from abroad — a trend confirmed by responses to our questionnaire. Concerning the other microwork platforms offering access interfaces in French, ClixSense, Clickworker, Appen and Ferpection are more visited than Wirk. This is both due to their open registration policy and to their lifespan (at least in the case of ClixSense, created as early as 2009). Some also benefit from their role as gateways to other microwork platforms: ClixSense points towards Figure Eight, while Clickworker provides access to UHRS (footnote \ref{UHRS}). To estimate the number of French microworkers on Mechanical Turk (right-hand column),  information from SimilarWeb.com was obviously not sufficient as the large number of observed unique visitors even exceeds the platform's own estimate of 500,000 (left column), and must therefore include surfers who are not themselves microworkers - not unlikely for a widely known platform. We thus rely on on a survey by the team led by Panos Ipeirotis at New York University\footnote{Data from the survey "Analyzing MTurk demographics" are available at \url{https://github.com/ipeirotis/mturk_demographics} and serve as the basis for the article by \cite{Difallah2018}.}\label{Ipeirotis}. Extrapolating from the size of the group of French "Turkers" that they observed, i.e. 0.25\% of the total, and assuming that their sample is representative, we estimate that the number of French residents operating on Mechanical Turk is around 1,250.

\subsection{Crossing methods to obtain a range of values}
\label{Crossing}

The calculation methods we have implemented so far provide  differing estimates of the population of microworkers in France: a high estimate is the number of signed-up users advertised by the platforms themselves (11,015,829 individuals worldwide), a low estimate is the plain capture-recapture model (2,792 individuals for a single platform, Wirk) and an intermediate estimate is the audience-based measure (62,754 individuals across platforms all over France).

Each of these approaches has limitations. At least in its basic, uncorrected version, the capture-recapture model underestimates the target population, while platforms' official declarations overestimate it. The use of audience measurements brings a helpful alternative perspective with intermediate results, but it is based on proprietary techniques whose actual implementation is difficult to audit for external researchers. At this stage, we have no means to choose one single metric from among those three, all the more so as they derive from very different measurement approaches. However, we can use these three results to cross-pollinate each other and help us derive more consistent measures.

Notice, first, that in the case of Wirk alone, the two corrections to the capture-recapture model that we have proposed to compensate for the downward bias of the basic Lincoln-Petersen formula (34,166 and 6,531), are close to, respectively, the official number of registered workers provided by the platform itself (50,000) and audience figures (6,958). These results implicitly validate all these metrics, although they are all lower (to different degrees) that the total number of platform-declared microworkers.

More importantly, audience figures enable an assessment of the number of registered microworkers who are located in France. To compute their number, we need to extrapolate their frequency of connection to the platforms (calculated from Table \ref{AudienceTab}) as a percentage of the total number of persons signed-up worldwide (Table \ref{DeclarativeTab})\footnote{Given the national scope of Wirk's activity, we consider all its registrations to be French. For Mechanical Turk we refer to the figures from P. Ipeirotis's survey mentioned above (footnote \ref{Ipeirotis}). For the other platforms, we take into account the number of visitors from France (Table \ref{AudienceTab}, last column) divided by the total number of visitors (Table \ref{AudienceTab}, first column), then apply the resulting percentage to the number of registrations (Table \ref{DeclarativeTab}). For example, Ferpection has 8,116 unique visitors from France, out of a total of 28,064 visitors, or 28.92\%; multiplying by its official number of registered users, equal to 50,000, we obtain 14,460.}. Table \ref{FrenchRegistrations} shows the results of these calculations.

\begin{table}%
\caption{Estimated number of people registered on microwork platforms and connecting from France.}
\label{FrenchRegistrations}
\begin{minipage}{\columnwidth}
\begin{center}
\begin{tabular}{ll}
  \toprule
  \textbf{Platform} & \textbf{Number French}\\
  & \\
  \hline
     & \\
   Amazon Mechanical Turk & 1,250 \\   
   Microworkers & 12,766 \\
   Clickworker & 72,720 \\
   ClixSense & 130,900 \\
   Wirk & 50,000 \\
   Ferpection & 14.460 \\
   Appen & 37,000 \\
  & \\
   \textbf{Total} & \textbf{319,096} \\
   
  \bottomrule
\end{tabular}
\end{center}
\bigskip\centering
\footnotesize\emph{Source:} Authors' elaboration.

\end{minipage}
\end{table}%

On this basis, we can go a step further and attempt to generalize the capture-recapture model initially used just on one platform, by applying its ratios (Table \ref{CaptureTab}, last column) to the total number of French microworkers we have just estimated (319,096). We thus obtain a range, with a low estimate of 17,869 (5.6\% of the declared total for France), a high estimate of 217,943 (with a "corrected" rate of 68.3\%) and an intermediate estimate of 41,482 (with the alternative correction of 13\%).

\subsection{Adjusting to take into account multi-homing}
\label{Multi-homing}

We cannot rely on these three estimates until we have made an additional correction to take into account multi-homing, that is, the presence of microworkers on several platforms at the same time. If we disregard multi-homing, we may over-estimate the number of microworkers by counting the same persons multiple times. First, how widespread is this phenomenon, or put differently, how many people does it concern? To answer this question, let us turn to our DiPLab survey, in which participants were asked to indicate their possible use of different microwork platforms from a list of options. We know that of our 908 unique respondents on Wirk, 151 are also registered on at least another microwork platform. In other words, 16.6\% of respondents practice multi-activity - the average usage being 1.27 platforms per person. 

Assuming our sample is representative, and extrapolating its behaviour to the entire French microworking population, amounts to applying the 16.6\% rate to the measures previously obtained (Table \ref{MultihomeTab}).

\begin{table}%
\caption{Estimate of the population of microworkers refined to take into account multi-homing.}
\label{MultihomeTab}
\begin{minipage}{\columnwidth}
\begin{center}
\begin{tabular}{p{1.9cm}p{2.3cm}p{1.7cm}p{1.7cm}}
  \toprule
  &  & \textbf{Result} & \textbf {Result}\\
  \textbf{Range} & \textbf{Measure} & \textbf{(reminder)} & \textbf{(minus multi-homing)} \\
  &  &  & \\
  \hline
  &  &  & \\
  Low & Uncorrected capture-recapture & 17,869 & 14,903 \\
  &  &  & \\
  Intermediate & Audience & 62,754 & 52,337 \\
  &  measure &  &  \\
  &  &  & \\
  High & Number of & 319,096 & 266,126 \\
  &  declared &  &  \\
  &  users (France) &  &  \\
  &  &  & \\
  \bottomrule
\end{tabular}
\end{center}
\bigskip\centering
\footnotesize\emph{Source:} Authors' elaboration.

\end{minipage}
\end{table}%

Little is known of the multi-homing practices of microworkers in other contexts but, comparing our survey to studies of other online practices such as the buying and selling of goods ans services, it appears that our rate of multi-activity is low. For example, \cite{Oxera2015} reports an average of 2.2 platforms per user. The DiPLab survey suggests that this relative smaller rate depends largely on the specific context of microwork. Many respondents hint that entry costs of multi-homing are relatively high, both in terms of incompatibility of payment systems (for example, Wirk uses Mango Pay while Microworkers uses Paypal, Skrill or Payoneer) and in terms of limited portability of "qualifications" (unpaid tests needed to access some types of micro-tasks). Additionally, the English language used on many international platforms puts off some French microworkers, even though some of their tasks can be in French.

\section{How to interpret results in light of microworkers' level of activity?}
\label{Interpretation}

We now turn to the meaning of these different measures of the size of the French microworking population. Using the DiPlab survey, we associate our figures to characterization of users according to their level of activity. We distinguish a group of "very active" (\ref{VeryActive}), a group of "routine" (\ref{Routine}) and one of "casual" (\ref{Casual}) microworkers.

\subsection{Very active microworkers}
\label{VeryActive}

Our first, lower estimate of 14,903 microworkers was initially calculated on the Wirk platform with the uncorrected capture-recapture model, subsequently extrapolated to the total number of French registrations, and finally corrected to remove duplicates. It can be interpreted as a measure of the number of users with a high level of activity. As discussed (section \ref{CaptureRecapture}), the number of captured and re-captured individuals in our survey is relatively higher than expected, and corresponds to a high level of activity: more engaged workers are significantly more likely to appear in both phases of the data collection. Another reason to believe that the great majority of the respondents to our survey fall into this category isthe frequency with which they log in to look for microtasks: at least once a week for 90\% of them.

Qualitative interviews conducted with a subset of respondents as part of the DiPLab project, confirm the hypothesis that there is a very active sub-population on Wirk and, by extension, on microwork platforms. This is the case for one respondent who, in addition to her primary occupation as a nurse, frequently performs "simple and quick tasks". Other users, expressing their "fear of missing out tasks", even go so far as leaving Wirk continuously open on their computers.

\subsection{Routine microworkers}
\label{Routine}

The measure we obtain through audience measurements from microwork platforms is much larger, with 52,337 users. It corresponds to a different, larger group of users, by definition those connected to a microwork platform (both Wirk and its competitors in France, as seen in section \ref{Audience}) at least once a month, after removing duplicates (section \ref{Multi-homing}).

Once again, DiPLab interviews provide us with insights into this mode of engagement with microwork. A respondent whose main job is editorial manager in a communication agency, has been microworking for three years. She signed up during a period of her life in which her "lifestyle" required "an additional income", resulting in "a compulsive need to earn more money". Since then, she has been returning to microwork "from time to time", with a preference for questionnaire-based micro-tasks that she sees as an extension of her education.

In sum, the group of "routine" users corresponds to a fragmented mode of engagement in microwork, with an often-selective attitude towards tasks.

\subsection{Casual microworkers}
\label{Casual}

Our last estimate features 266,126 persons and corresponds to registered users as declared by platforms themselves, after removing non-French users and duplicates. It is a group of "casual" microworkers : French users who signed up with one or more microwork platforms, who may or may not have an intense level of activity. Some of them may perform less than one task per month. 

The long tail of this population that connects less frequently, is more difficult to reach through surveys and may escape audience measures. Yet our interviews provide sufficient insight to portray it. We can distinguish two sub-populations. One includes users who signed up out of curiosity and quickly abandoned microwork because of the scarcity and difficulty of the tasks - two points often emphasized by our study participants. Another includes individuals who were "involved in" microwork at least once and who can, intermittently or in the future, become "routine" or even "very active" users.

One respondent, holding a day-job as administrative secretary, illustrates this situation well. Registered on Wirk in 2016, he "quickly dropped out". A few months later, he received an email indicating that the platform would delete his account due to inactivity. He then decided to resume his microwork in a relatively sustained way. He now claims he earns about 80 euros a month from micro-tasks, almost four times more than the average in our survey. Another microworker, currently seeking employment, also alternates between inactivity and intensive performance. In her working life, she has always experienced alternating periods of unemployment and periods of employment. She says she stops microwork as soon as she finds a job because she will "no longer have time", while her phases of unemployment correspond to periods of intensive activity on Wirk because she prefers to "do this rather than nothing".

Table \ref{Results} summarizes our results. 

\begin{table}%
\caption{Number of microworkers in France, by level of activity.}
\label{Results}
\begin{minipage}{\columnwidth}
\begin{center}
\begin{tabular}{p{1.5cm}p{2.5cm}p{1.7cm}p{1cm}}
  \toprule
  \textbf{Group} & \textbf{Type of activity} & \textbf{Frequency} & \textbf{Size} \\
  &  &  & \\
  \hline
  &  &  & \\
  Very active & Simple and quick micro-tasks & At least once a week & 14,903 \\
    &  &  & \\
  Routine & Selected micro-tasks & At least once a month & 52,337 \\
    &  &  & \\
  Casual & Alternating periods of activity and inactivity & Less than once a month & 266,126 \\
  &  &  & \\
  \bottomrule
\end{tabular}
\end{center}
\bigskip\centering
\footnotesize\emph{Source:} Authors' elaboration.

\end{minipage}
\end{table}%
\section{Conclusions}
\label{Conclusions}

With this article, we have sought to launch a discussion on how to quantify the pervasiveness of new forms of work that, for the time being, are largely outside official statistics. Microwork illustrates this difficulty very clearly. Lacking the media coverage that platforms such as Uber or AirBnb receive, microwork is largely unrecognised, overshadowed by the commercial propaganda revolving around the promises of algorithms and robotic exploits \cite{Irani2015}. The voice of the individuals who perform microwork goes largely unheard. Somewhat paradoxically, microwork remains at the margin of formal economy even though it constitutes a strategic resource for big data-intensive artificial intelligence, a major component of the digital economy.

We have proposed an estimate of the size of the microworking population in France. We have chosen to take into account not just one platform (as most studies that have exclusively focused on Amazon mechanical Turk so far) but a set of platforms that we have reasons to consider relevant in the country. We have combined different methods and data sources: from the public declarations of platforms themselves and the results of publicly-available audience measurement panels, to data that we collected through questionnaires and interviews as part of our DiPLab project. In this way, we have detected and qualified three groups of users according to their level of involvement: 14,903 "very active" individuals, most of whom are present on the platforms at least once a week; 52,337 "routine" users, present at least once a month, with a more fragmented level of activity and a selective choice of tasks; and 266.126 "casual" users who mostly alternate between inactivity and periods of more intensive practice of microwork, notably during phases of transition in their lives (unemployment, maternity leave, etc.). 

We mentioned the limitations of the publicly available information we have used, including the opacity of audience measures - constructed by private companies for their clients and difficult to audit for independent researchers. In addition as we said, our DiPLab survey included just two waves of capture and recapture, allowing only for approximate corrections of the bias resulting from the equal-capture-chances assumption. Another limitation to our study lies in the difficulty of assessing the extent to which these three sets of microworkers, corresponding to three platform usage patterns and established through three very different estimation methods, intersect so that their populations partly overlap. Despite the questions they still raise, these figures provide extremely useful orders of magnitude. 

Consequently, our estimates are critical contributions to the understanding of ongoing significant changes in the labor market in the era of digital platforms. The tens of thousands of French microworkers performing increasingly fragmented piece-based tasks can have a strong impact on the very perception of labor, its social protection and its modes and levels of remuneration in the coming years.

By way of comparison, the IGAS report we already mentioned (section \ref{Declarative}) estimated the number of Uber drivers in France at around 14,000 \cite{AmarViossat2016}; all platforms combined, there are around 27,000 drivers of private hire vehicles in France, according to the Ministry of transport (cited in \cite{Pommier2018}). As for bicycle delivery companies, Deliveroo claims to have 9,300 couriers in France \cite{Deliveroo2018}. Our estimates therefore shows that microwork has a workforce whose size is at least comparable to that of these high-profile platforms - if not higher if we factor in casual participation. It is therefore not an anecdotal phenomenon and deserves great attention from the research community, from unions and from policy-makers alike.

In France and in Europe, the current debate on social protection for digital platforms workers must take microwork into account. Research must sustain these reflections and provide elements for action, by proposing a more accurate knowledge of the social changes that come with microwork as well as of the operational functioning and business models of the digital platforms that benefit from it.

\begin{acks}

The authors would like to thank current and former DiPLab team members, notably Maxime Besenval, Odile Chagny, Marion Coville, Touhfat Mouhtare, Lise Mounier, and Manisha Venkat, for their inputs to this paper and suggestions to improve it. Any remaining errors or inconsistencies are exclusively ours.

The DiPLab research project is co-funded by Maison des Sciences de l'Homme Paris-Saclay; Force Ouvri\`ere, a workers' union; and France Strat\'egie, a service of the French Prime Minister's office. We also thank Wirk for logistical and financial support when fielding our survey, and Inria for complementary funding. For more information: http://diplab.eu/

\end{acks}

\bibliographystyle{ACM-Reference-Format}
\bibliography{LeLudec_Tubaro_Casilli_05012019b.bib}

\end{document}